\documentclass[oribibl]{llncs}
\usepackage{amsmath}
\usepackage{color}
\usepackage{amssymb,xspace,latexsym,epic,eepic,graphicx}

\usepackage{hhline}
\usepackage{amssymb}
\usepackage{latexsym}
\usepackage{times}
\usepackage[lflt]{floatflt}
\usepackage{hyperref}

\newcommand{\boxtheorem}{\hfill $\blacksquare$}
\newcommand{\ignore}[1]{}

\newcommand{\nit}[1]{{\it #1}}
\newcommand{\n}{~{\it not}~}

\newcommand{\tr}{\mathbf{t}}

\newcommand{\ta}{\mathbf{t}}
\newcommand{\fa}{\mathbf{f}}

\newcommand{\tss}{{\bf \tr^{\star\star}}}
\newcommand{\IC}{\nit{IC}}

\newcommand{\Q}{{\cal Q}}
\newcommand{\A}{\nit{Ans}}

\newcommand{\f}{{\bf f}}
\newcommand{\s}{\!_{\star}}
\newcommand{\ds}{\!_{\star\star}}

\newcommand{\p}{P\!_f}

\newcommand{\R}{{\cal R}}
\newcommand{\F}{\nit{FD}}

\title{{\bf Second-Order Specifications and Quantifier Elimination for  Consistent Query Answering in Databases\vspace{-4mm}}\thanks{This is a slightly  extended and updated version of \cite{arequipa}, and of a forthcoming extended abstract based on the latter.}}

\author{{\bf Leopoldo Bertossi}\\
Universidad Adolfo Ib\'a\~nez\\
Faculty of Engineering and Sciences \\
and\\
Millennium Institute for Foundational Research on Data (IMFD)\\
Santiago, Chile\\
leopoldo.bertossi@uai.cl \vspace{-4mm}}
 \institute{}
\date{}

\begin{document}

\pagenumbering{arabic}
\pagestyle{plain}
\bibliographystyle{plain}

\maketitle

\begin{abstract}
Consistent answers to a query  from a possibly inconsistent database are answers
that are simultaneously  retrieved from every possible repair of the database. Repairs
are consistent instances that minimally differ from the original inconsistent instance. It has been shown before
that database repairs can be specified as the stable models of a disjunctive logic program. In this
paper we show how to use the repair programs to transform the  problem of consistent query answering
into a problem of reasoning w.r.t. a theory written in second-order predicate logic. It also investigated how a first-order
theory can be obtained instead by applying second-order quantifier elimination techniques.
\end{abstract}

%%%%%%%%%%%%%%%%%%%%%%%%%%%%%%%%%%%%%%%%%%%%%%%%%%%%%%%%%%%%%%%%%%%%%%%%%%
\section{Introduction}\label{se:introduction}\vspace{-3mm}

Integrity constraints (ICs) are conditions that come with a relational database schema ${\cal S}$, and are expected to
be satisfied by the instances of ${\cal S}$. In this way,  database instances stay in
correspondence with the outside reality they intend to model. If an instance of $\cal S$
does not satisfy the ICs, it is said to be {\em inconsistent}. For several reasons a database instance
may become inconsistent, and in consequence, it is only partially semantically correct.

{\em Consistent query answering} (CQA) in databases is an area of data management that attempts to  characterize and compute
answers to a query that are consistent with respect to (w.r.t.)  a given set of ICs. These
problems are relevant because the database instance being queried possibly fails to satisfy the set of ICs as a whole.
So, only locally consistent information is expected to be extracted from the database. These problems have been
investigated by the database community at least since the notion of consistent query answer was explicitly
introduced in \cite{ABC1999}. Cf. \cite{B2006,chom07,bertossi2011} for surveys of  CQA.

Informally, a tuple of constants  $\bar{t}$ is a consistent answer from an instance $D$ to a query ${\cal Q}(\bar{x})$ w.r.t. a set of
ICs $\IC$ if $\bar{t}$ can be
obtained as a usual answer to $\cal{Q}$ from every {\em repair} of $D$. Here, a {\em repair} is a consistent instance for the schema ${\cal S}$
that differs from $D$ by a minimal set of database atoms under set inclusion \cite{ABC1999}.

In \cite{BB2003} it was shown how repairs of a database $D$ w.r.t. a set of ICs can be specified as the {\em stable models of a disjunctive logic program} \cite{GL1991}
$\Pi$, by a so-called {\em repair program}, whose
set of facts corresponds to the original instance $D$. In this way, obtaining consistent answers  becomes a problem of reasoning over the class
of all stable models of $\Pi$. More loosely, logic programs with stable model semantics are also called {\em answer-set programs} \cite{GL1991}, and their stable models are their answer sets. Answer-set  programming  has become a paradigm and  powerful tool for the specification and solution of hard combinatorial problems \cite{brewka}.

\begin{example} \label{ex:intro}
Consider a relational database schema $\cal S$ with a predicate $P(X,Y)$
and the functional dependency (FD)~ $X \to Y$, stating that the first attribute functionally determines
the second. It can be expressed in $L({\cal S})$, the first-order (FO) language of predicate logic associated to $\cal S$, as the sentence $\forall x \forall y \forall z(P(x,y) \wedge
P(x,z) \rightarrow y = z)$. $D = \{P(a,b),
P(a,c), P(d,e) \}$ is an inconsistent instance: the first
two tuples jointly violate the FD.  The instance has two repairs: \
$D_1 = \{P(a,b),
P(d,e) \}$ and $D_2 = \{
P(a,c), P(d,e) \}$. The query ${\cal Q}_1(y)\!:  \exists x P(x,y)$ has only one consistent
answer,  the tuple $(e)$, whereas the query   ${\cal Q}_2(x)\!:  \exists y P(x,y)$
has $(a), (d)$ as consistent answers.  This is because they are standard answers to the queries from both repairs.

The
repairs can be specified as the stable models of a logic program that contains, among other rules,
a main rule that takes care of restoring consistency w.r.t. the FD, namely  a rule of the form
$$P(x,y,{\bf f}) \vee P(x,z,{\bf f})  \leftarrow P(x,y),
P(x,z), y \neq z.$$It specifies that whenever the FD is violated, which is captured by the body of the
rule, then one (and only one if possible) of the two tuples involved in the violation has to be deleted (made false, as indicated by the annotation constant
$\mathbf{f}$), which is
captured by the disjunctive head.

Repair programs can always be used for CQA. However, as shown in \cite{ABC1999}, it is sometimes possible to obtain CQA by posing a
new FO query to the inconsistent database, which is much simpler to do. For example, the consistent answers to the query ${\cal Q}_3\!: P(x,y)$ can be obtained rewriting the query
into ${\cal Q}_3^\prime\!: P(x,y) \wedge \neg \exists z(P(x,z) \wedge z \neq y)$, and posing it to the original instance $D$. That is, $(t_1,t_2)$ is a consistent
answer to ${\cal Q}_3$ iff $D \models {\cal Q}_3^\prime[t_1,t_2]$. \boxtheorem
\end{example}
In an ideal situation, consistent answers to a query ${\cal
Q}$ from a database instance $D$ should be obtained by posing a
new query ${\cal Q}'$ to $D$, as an ordinary query that is, hopefully, easy to evaluate against $D$. This is the case,
for example, when
 ${\cal Q}'$ is a query expressed in the first-order (FO) language $L({\cal S})$ associated to the schema ${\cal S}$, as in the example we just saw.

Some classes of queries and ICs with this property have been already identified \cite{ABC1999,CM2005,FM2007}. Many more cases have been identified by Wijsen in a series of papers (c.f. \cite{wijsenSurvey} and \cite{wijsenSigRec} for excellent surveys).  See \cite{Paraschos20}  for much more recent results in relation to a full classification of the complexity of CQA under {\em key constraints}. That is, under FDs that determine all the attributes in a predicate \cite{AHV1995}.

The main point is that, at least under  key constraints,  one can syntactically classify and decide conjunctive queries (CQs) in terms of their data complexity for consistent query answering.\footnote{As usual in databases, all the complexity results in this
paper are about {\em data complexity}, i.e. in terms of the size of the database instance.} A trichotomy appears: a CQ can be FO-rewritable, or in \nit{PTIME} ($L$-complete), or \nit{coNP}-complete. There are queries for these three classes. For the first class, the rewriting can be computed.
\ In this case, of course, it is
possible to compute the consistent answers to ${\cal Q}$ in
polynomial time in the size of $D$.

It is worth emphasizing that there are CQs for which consistent query answering can be done in polynomial time, but provably not via FO-rewriting \cite{wijsenIS09,wijsenIPL10}. This opens the question about the right logical language for a rewriting, if any.

\ignore{Unfortunately, the
methodology for CQA based on FO query rewriting is bound to
have limited applicability, for CQA can have a higher data
complexity.  Even for a conjunctive
query ${\cal Q}$ and certain functional dependencies, CQA can be
$\nit{coNP}$-complete \cite{CM2005,FM2007}.}

Repair programs provide a general mechanism for computing
consistent answers.
Actually, the data complexity of CQA can be as high as the data complexity of cautious query evaluation from disjunctive logic
programs under the stable model semantics, namely $\Pi^P_2$-complete \cite{DEGV2001,CM2005}.  However, assuming that the
polynomial hierarchy does not collapse, this may be an expensive mechanism for queries
that can be answered more efficiently, e.g. in polynomial time, as is the case for
some classes of queries and
ICs. It turns out that the complexity landscape between FO rewritable cases and
$\Pi^P_2$-completeness for CQA is still not quite clear.
 Results obtained in the middle ground are scattered, isolated, and rather ad hoc.

 In those cases where a FO rewriting for CQA is possible, one can  transform the problem into  one of
reasoning in classical predicate logic (cf. Example \ref{ex:intro}), because the original database can be ``logically reconstructed" as a FO theory
\cite{R1984}. In this work we investigate how repair programs can be used to generate a theory written in classical
logic from which CQA can be captured as logical entailment. This theory can be written in second-order or first-order predicate logic. We start by trying to achieve the former.
For this, we provide concrete  specifications of database repairs
in second-order (SO) predicate logic. They are obtained by applying recent
results on the specification in SO logic of the stable models of a logic
 program \cite{FLL2007,ExtFerraris} -in our case, the repair programs- and
older results on their characterization as the models of a circumscription theory \cite{M1980} in the case of disjunctive
stratified programs \cite{przy88,przy91}. This circumscription can be specified in SO predicate logic \cite{Lifschitz1985}.

In order to achieve a first-order specification, for some cases related to queries and functional dependencies (FDs), we apply the techniques for SO quantifier elimination that have been
introduced and studied in \cite{DLS1997}. In this way it is possible to obtain a FO specification of the database repairs.
This transforms the problem of CQA into a problem of logical reasoning in FO logic. We illustrate by means of an example
how to obtain a FO rewriting for CQA from this specification. Actually, in this work we will  concentrate mostly on FDs, and
key constraints in particular. We concentrate mostly on a detailed example that illustrates the technique for SO quantifier elimination. Generalizing the methodology to more general cases is left for future investigation.
C.f. Section \ref{sec:todo}), where we also discuss the possibility of obtaining rewritings in fixed-point logic, when it is provably the case that no FO rewriting exists.

Most of the complexity results
in CQA have been obtained for this class of constraints, but the complexity is not fully understood yet.
We expect that the kind of results obtained in this work will help shed more light on this picture, in particular with respect
to (w.r.t.) rewritability for CQA. These applications and others, like a better understanding of ``the logic of CQA", are still to be developed.

This paper is structured as follows.  In Section \ref{sec:frame} we introduce background material, illustrating  with examples some fundamental notions and
constructions.  In Section \ref{sec:SO} we obtain second-order specifications of database repairs from database repair programs. In Section
\ref{sec:QE} we concentrate on functional dependencies, applying quantifier elimination techniques to obtain first-order specifications
of repairs. In Section \ref{sec:todo}, we speculate about the possibility of obtaining query rewritings in fixed-point logic for cases where FO rewriting provably do not exist. In Section \ref{sec:concl} we draw some conclusions; we point to ongoing and future research, and also to some promising
research directions that are  opened by our research.

\vspace{-3mm}
\section{Background and Preliminaries} \label{sec:frame}%\vspace{-2mm}

\vspace{-2mm}
\paragraph{\bf \em Relational databases.} \
Let ${\cal S}$ be a relational schema. It contains a possible
infinite domain $\mathcal{U}$ and a set of predicates. $\cal S$
determines a language $L({\cal S})$ of first-order predicate
logic, in which the elements of $\mathcal{U}$ appear as
constants. Integrity constraints are sentences in this language that are expected to be satisfied
by a database instance. In order to simplify the presentation, we assume in this work that they are
universal sentences, because existential ICs, like
referential integrity constraints, require a slightly different treatment in the context of CQA
 \cite{BB2006}. For the same reason, we assume that database instances do not have null values \cite{BB2006}.

 A database instance $D$ for $\cal S$ is a finite set
of ground atoms of the form $R(\bar{a})$, with $R \in {\cal S}$
and $\bar{a}$ is a tuple of constants in
$\mathcal{U}$.\footnote{When we write something like $R \in {\cal S}$, we understand that $R$ is a database predicate, not a built-in. For a tuple of constants $\bar{a} =
(a_1, \ldots, a_k)$, $\bar{a} \in \mathcal{U}$ denotes $a_i \in
\mathcal{U}$ for $i = 1, \ldots, k$.} $D$ can also be seen as a
Herbrand structure \cite{L1987} for interpreting $L({\cal S})$,
namely $\langle \mathcal{U}, (R^D)_{R \in {\cal S}}, (u)_{u \in
\mathcal{U}}\rangle$, with $R^D = \{R(\bar{a})~|~R(\bar{a}) \in
D\}$. The database $D$ can also be logically reconstructed
as a first-order sentence ${\cal R}(D)$ as done by Reiter in
\cite{R1984}.

\begin{example} \label{ex:recons} If ${\cal S}$ has the domain $\mathcal{U} = \{a,b,c,d,e,f,g\}$
and only predicate $P(\cdot,\cdot)$, then $D = \{P(a,b), P(a,c),
P(d,e)  \}$   could be a database instance for $\cal S$.

In this case, $\R(D)$ is the conjunction of the following
sentences: (a) Domain Closure Axiom (DCA):~ $\forall x(x = a \vee x
= b \vee x =c \vee x = d \vee x =e \vee x = f \vee x = g)$. (b)
Unique Names Axiom (UNA):~ $(a \neq b \wedge \cdots \wedge f \neq
g)$. (c) Predicate Completion (PC):~ $\forall x \forall y(P(x, y)
\equiv (x=a \wedge y= b) \vee (x = a \wedge y =c) \vee (x=d
\wedge y =e))$. The theory $\R(D)$ is categorical, i.e. $D$ is
essentially its only model, which is, also essentially, a Herbrand model.
\boxtheorem
\end{example}
In the previous example, the domain is finite, which makes it possible
to use a domain closure axiom. If the domain $\mathcal{U}$ is
infinite, the domain closure axiom (DC) is applied to the active domain, $\nit{Ac}(D)$, of the database, i.e. to the set
of constants appearing in the relations of the database instance. Since the extensions of
the predicates are always finite, we can always build a DC axiom and predicate
completion axioms for them. For static databases and CQA w.r.t. universal ICs, the active domain of the original database is large enough to
restore consistency, and in particular, to define the repairs. If we restrict ourselves to
Herbrand structures, we do not need the domain closure or the
unique names axioms.

In this work we consider database queries $\Q$ that are safe and  written in the FO language $L({\cal S})$. Most frequently, we consider
conjunctive query with built-ins  \cite{AHV1995}.

\vspace{-2mm}
\paragraph{\bf \em Answer-set programs.}
\ We consider disjunctive Datalog programs $\Pi$ \cite{eiterGottlob97} with a finite number of rules of the form
$$A_1 \vee \ldots A_n \leftarrow P_1, \ldots, P_m, \n N_1, \ldots, \n N_k,$$
with $0\leq n,m,k$, and the $A_i, P_j, N_s$ are positive FO atoms. The terms in these atoms are constants or variables. The variables in the $A_i, N_s$ appear all among those
in the $P_j$.   The constants in the program $\Pi$ form the (finite) Herbrand universe $U$ of the program. The ground version of
program $\Pi$, $\nit{gr}(\Pi)$, is obtained by instantiating the variables in $\Pi$ in all possible combinations  using
values from $U$. The Herbrand base $\nit{HB}$ of $\Pi$ consists of all the possible atomic sentences obtained by instantiating the
predicates in $\Pi$ in $U$. A subset $M$ of $\nit{HB}$ is a model of $\Pi$ it is satisfies $\nit{gr}(\Pi)$, that is: For every
ground rule $A_1 \vee \ldots A_n \leftarrow P_1, \ldots, P_m, \n N_1, \ldots,
\n N_k$ of $\nit{gr}(\Pi)$, if $\{P_1, \ldots, P_m\} \subseteq M$ and $\{N_1, \ldots, N_k\} \cap M = \emptyset$, then
$\{A_1, \ldots, A_n\} \cap M \neq \emptyset$. $M$ is a minimal model of $\Pi$ if it is a model of $\Pi$, and $\Pi$ has no model
that is properly contained in $M$. $\nit{MM}(\Pi)$ denotes the class of minimal models of $\Pi$.

Now, take $S \subseteq \nit{HB}(\Pi)$, and transform $\nit{gr}(\Pi)$ into a new, positive program $\nit{gr}(\Pi)\!\downarrow$ (i.e. without $\nit{not}$), as follows:
Delete every rule  $A_1 \vee \ldots A_n \leftarrow P_1, \ldots,P_m, \n N_1,$ $ \ldots,
\n N_k$ for which $\{N_1, \ldots, N_k\} \cap S \neq \emptyset$. Next, transform each remaining rule $A_1 \vee \ldots A_n \leftarrow P_1, \ldots, P_m, \n N_1, \ldots,
\n N_k$ into $A_1 \vee \ldots A_n \leftarrow P_1, \ldots, P_m$. Now, $S$ is a stable model of $\Pi$ if $S \in \nit{MM}(\nit{gr}(\Pi)\!\downarrow)$.

A disjunctive Datalog program  is {\em stratified} if its set of predicates ${\cal P}$ can be partitioned into
a sequence ${\cal P}_1, \ldots, {\cal P}_k$ in such a way that, for every  $P \in {\cal P}$:
\begin{enumerate}
\item If $P \in {\cal P}_i$ and predicate $Q$ appears in a head of a rule
with $P$, then $Q \in {\cal P}_i$.
\item If $P \in {\cal P}_i$ and $Q$ appears positively in the body of a rule that has $P$ in the head, then $Q
\in {\cal P}_{\!j}$, with $j \leq i$.
\item If $P \in {\cal P}_i$ and $Q$ appears negatively in the body of a rule that has $P$ in the head, then $Q
\in {\cal P}_{\!j}$, with $j < i$.
\end{enumerate}
If a program is stratified, then its stable models can be computed bottom-up by propagating data upwards from the
underlying extensional database, and making sure to minimize the selection of true atoms from the disjunctive heads.
Since the latter introduce a form of non-determinism, a program may have several stable models.

\vspace{-2mm}
\paragraph{\bf \em Database repairs and repair-programs.} \ %\label{sec:rep}
The following definitions were introduced in \cite{ABC1999}: A {\em repair} of  instance $D$ w.r.t. a set \IC~ of ICs is an
instance $D^\prime$, over the same schema ${\cal S}$, that  satisfies
${\it IC}$, i.e. $D'
\models \IC$, and makes the symmetric set-difference $\Delta(D,D')$~ minimal w.r.t. set inclusion. $\nit{Rep}(D,\IC)$ denotes
the set of repairs of $D$ w.r.t. \IC.

Given a database instance $D$ and a set of ICs, \IC, a disjunctive
answer set program \cite{GL1991}, called a {\em repair program},
can be used to specify the repairs of $D$ w.r.t. $\IC$. More precisely,
(all and only) the  repairs of $D$ can be read-off from the stable models of the
program.
 Because of their simplicity and scope, we will use the repair programs first introduced in \cite{BB2003} in their slightly modified version in
\cite{monicaJournal08}. The most general version, that can be used for restoring consistency w.r.t. to ICs that include
referential constraints and database instances with null values can be found \cite{BB2006,monicaJournal08}. However, in this paper we do
not consider null values. Other earlier forms of repair programs can also be found in \cite{ABC2003,GGZ2003}.

Repair programs use ~annotation constants  in an~ extra
argument of each of ~the database predicates. More precisely,
for each $n$-ary $P \in {\cal S}$, we make a copy $P\!\!\_$ ,
which is $(n\!+\!1)$-ary.  The intended semantics of the
annotations is given in Table \ref{tab:annot}.

\vspace{-2mm}
\begin{table}[h]
\begin{center}\begin{tabular}{|l|l|l|}
\hline
Annotation &  Atom & The tuple $P(\bar{a})$ is:\\
\hline
$\ta$ & $P\!\!\_(\bar{a},\ta)$ & ~made true/inserted\\
$\fa$ & $P\!\!\_(\bar{a},\fa)$ & ~made false/deleted\\
$\tr^\star$ & $P\!\!\_(\bar{a},\tr^\star)$ &  ~true or made true\\
%$\fs$ & $P\!\!\_(\bar{a},\fs)$ &  ~false or made false\\
$\tss$ & $P\!\!\_(\bar{a},\tss)$ &  ~true in the repair\\
\hline
\end{tabular}
\vspace{2mm}\caption{Annotations with Intended Semantics}\label{tab:annot}
\end{center}
\end{table}

\vspace*{-1.3cm}
\begin{example} \label{ex:genProg} Consider $IC\!:~
\forall xy (P(x,y) \rightarrow Q(x,y))$; and the inconsistent database instance
$D =\{P(c,l), P(d,m),
Q(d,m), Q(e,k) \}$. The repair program $\Pi(D,{\it IC})$ has the following rules (and facts):
\begin{enumerate}
\item [1.] ~ Original database facts: ~~$P(c,l),$ etc.

\item [2.] Whatever was true or becomes true, is
annotated with ~$\mathbf{t^\star}$:

$P\!\!\_(\bar{x},\mathbf{t^{\star}}) \leftarrow
P(\bar{x}).$~~~~ $P\!\!\_(\bar{x},\mathbf{t^{\star}}) \leftarrow
P\!\!\_(\bar{x},\mathbf{t}).$ ~~~~(the same for $Q$)

\item [3.] There may be interacting ICs (not here), and the
repair process may take several steps, changes could trigger other
changes:

$P\!\!\_(\bar{x},\mathbf{f})~ \vee ~Q\!\!\_(\bar{x},\mathbf{t})
~\leftarrow~
P\!\!\_(\bar{x},\mathbf{t^{\star}}), Q\!\!\_(\bar{x},\mathbf{\f}).$

$P\!\!\_(\bar{x},\mathbf{f})~ \vee ~Q\!\!\_(\bar{x},\mathbf{t})
~\leftarrow~
P\!\!\_(\bar{x},\mathbf{t^{\star}}), \n Q(\bar{x}).$

Two rules per IC that say how to repair the satisfaction of the IC (cf. the head) in
case of a
violation (cf. the body).
Passing to annotation $\mathbf{t^{\star}}$  allows to keep repairing the database w.r.t.
to all the ICs until the process stabilizes.

\item [4.]  Program
 constraints: ~~~~
$\leftarrow P\!\!\_(\bar{x},\mathbf{t}),
P\!\!\_(\bar{x},\mathbf{f}).$ ~~~~ (similarly for $Q$)

\item [5.] Annotations constants
$\mathbf{t^{\star\star}}$ are used to read off the
atoms in a repair:

$P\!\!\_(\bar{x},\mathbf{t^{\star\star}}) \leftarrow
P\!\!\_(\bar{x},\mathbf{t}).$~~~~
$P\!\!\_(\bar{x},\mathbf{t^{\star\star}}) \leftarrow
P(\bar{x}),~\mathit{not}~P\!\!\_(\bar{x},\mathbf{f}).$~~~~ (similarly for $Q$)
\end{enumerate}
The {\em program constraints} in Item 4. are used to filter out {\em incoherent models} where a tuple is both inserted and deleted. In this particular
 example, we actually do not need program denials, because a tuple can never be both deleted and
inserted. However, we keep them for illustration purposes; they may be necessary when there are interacting
ICs \cite{monicaJournal08}.\footnote{In Appendix \ref{sec:app1}, we give the general program for any set of universal ICs and databases without null values.} \boxtheorem
\end{example}
For simplicity,  from now on, we use $P\!_f(\_,\_)$ for
$P\!\!\_(\_,\_,\f)$, $P\ds(\_,\_)$ for $P\!\!\_(\_,\_,\tss)$,
etc. That is, annotations are replaced by new predicates.
 The repairs are in one-to-one correspondence with the restriction of the stable models to
their atoms annotated with $\tss$ (or to predicates of the form $P\ds$) \cite{BB2006}.

\vspace{-2mm}
\paragraph{\bf \em Queries and consistent answers.} \ %\label{sec:cqa}
For a query $\mathcal{Q}(\bar{x}) \in
L(\mathcal{S})$, we say that $\bar{a} \in
\mathcal{U}$ is a {\em consistent answer to} $\mathcal{Q}$ in
$D$ w.r.t. $\IC$, denoted $D \models_c \mathcal{Q}(\bar{a})$, iff
$D' \models \mathcal{Q}(\bar{a})$ for every $D' \in
\nit{Rep}(D,\IC)$.

In this paper, the query $\Q$ above is a safe FO query written in the language $L({\cal S})$ \cite{AHV1995}.
 In order to pose this query to the models of the repair program, i.e.
to the repairs, the query has to be reformulated as a query $\Q^{\star\!\star}$ that is obtained from $Q$ by replacing each database
predicate $P$ by its double starred version $P\ds$. For example, for $\Q(y)\!: \exists x (P(x,y) \wedge \neg Q(x,y) \wedge x \neq y)$, we have
$\Q^{\star\!\star}(y)\!: \exists x (P\ds(x,y) \wedge \neg Q\ds(x,y) \wedge x \neq y)$.
The query $\Q$ could also be written in (safe) Datalog  (or in any of its extensions) \cite{AHV1995}. In this case, $\Q^{\star\!\star}$ is
obtained from $\Q$ by replacing every extensional predicate $P$ (in ${\cal S}$) by $P\ds$.

The repair programs can also be used to obtain
consistent answers to queries, as
the  cautions or skeptical answers from
the combined program consisting of the repair program and a
query program.
So, given a FO query  ${\cal Q}(\bar{x})$,  $\Q^{\star\!\star}(\bar{x})$ is rewritten as a Datalog
query $\Pi^\Q$, possibly containing weak negation, $\n\!\!$. ~$\Pi^\Q$ contains a predicate, $\nit{Ans}^\Q(\bar{x})$, to collect
the final query answers.
If the query $\Q$ is given directly as a Datalog program with negation, then $\Pi^Q$ is simply $\Q^{\star\!\star}$.
To simplify things on the query side, and according to the usual conventions, we will assume that such Datalog queries $\Q$
are stratified normal programs, most usually, a non-recursive Datalog\!\!$^{\n}$ query \cite{AHV1995}, that is obtained as
a translation of a FO query.

 In order to obtain the consistent answers to $\Q$, the query program $\Pi^\Q$ is combined with the
repair program $\Pi(D,\IC)$ into a new program $\Pi$.
The extension of the answer predicate $\A^{\!\Q}$  in the intersection of all stable models of $\Pi$ contains exactly
the consistent answers.
That is, it holds
\begin{eqnarray}
D \models_{c} {\cal Q}(\bar{a}) ~&\Longleftrightarrow&~ D' \models {\cal Q}(\bar{a}), \mbox{ for every } D' \in
\nit{Rep}(D,\IC) \label{eq:cqaLP}\\
~&\Longleftrightarrow&~ \Pi(D,\IC) \cup \Pi^\Q
\models_\nit{cs} \nit{Ans}^{\!\Q}(\bar{a}),
\end{eqnarray}
where  $\models_{\!{\it cs}}$
stands for {\em cautious}, i.e. being true in all
stable models of $\Pi$.

If on the LHS of (\ref{eq:cqaLP}), $\Q$  is already a Datalog program, $D' \models {\cal Q}(\bar{a})$ means that
$\bar{a}$ is an answer to the Datalog query when using $D'$ as the underlying extensional database of program facts.

\begin{example} (example \ref{ex:genProg} continued) The query $\Q(x):~ \exists y (P(x,y) \wedge \neg Q(x,y)) \vee  \exists y (Q(x,y) \wedge P(x,y))$ can
also be written as the non-recursive
Datalog$^\nit{not}$ query
\begin{eqnarray*}
\A(x) &\leftarrow& P(x,y), \n Q(x,y)\\
\A(x) &\leftarrow& Q(x,y),  P(x,y).
\end{eqnarray*}
The corresponding query program $\Pi^{\!\Q}$ for consistent query answering is
\begin{eqnarray*}
\A(x) &\leftarrow& P\ds(x,y), \n Q\ds(x,y)\\
\A(x) &\leftarrow& Q\ds(x,y),  P\ds(x,y).
\end{eqnarray*}
It holds~ $D \models_c {\cal Q}(a_1,a_2)$ iff $(a_1,a_2) \in \A^M$ for every stable model
$M$ of $\Pi(D,\IC) \cup \Pi^{\!\Q}$. Here, $\A^M$ is the extension of predicate $\A$ in $M$.
\boxtheorem
\end{example}

\vspace{-2mm}
\paragraph{\bf \em Functional dependencies and repairs.} \
In this work we will mostly concentrate on functional dependencies (FDs), and key constraints (KCs), in particular.
For some classes of KCs and conjunctive queries there are efficient
algorithms for  CQA based on FO query rewriting \cite{ABC1999,CM2005,FM2007,wijsenSurvey}. In \cite{wijsenIS09,wijsenIPL10} there are examples
of conjunctive queries for which CQA w.r.t.  KCs is in $\nit{PTIME}$,
but there is  no consistent FO rewriting of the query.
FDs are particular cases of {\em denial constraints}, i.e. ICs of the form $\bar{\forall} \neg (A_1 \wedge
\cdots \wedge A_m)$, where the $A_i$ are database or built-in atoms, and $\bar{\forall}$ denotes the universal closure of the formula.

In \cite{BBB2003}, it is proved that for certain classes of ICs, that include all denial constraints, the repair programs
become {\em head-cycle free} (HCF). For them cautious query evaluation becomes ${\it coNP}$-complete \cite{BenDechter94,DEGV2001}.
Thus, we obtain that CQA
of conjunctive queries  w.r.t. functional dependencies belongs $\nit{coNP}$ \cite{BBB2003}.
For conjunctive queries and certain functional dependencies (actually, a single key dependency suffices),
CQA turns out to be $\nit{coNP}$-complete \cite{CM2005,FM2007,wijsenSurvey}, matching the general upper bound provided by the repair program.

If for a relational
predicate $R$, if we have the FD $\bar{X} \rightarrow Y$, where $\bar{X}$ is a set of attributes $\{X_1, \ldots, X_n\}$ and $Y$ a single attribute,
the repair program contains a rule of the form
\begin{eqnarray}
R_{\!f}(& \hspace{-7mm}\ldots,x_1, \ldots, x_n, \ldots, y_1, \ldots)  \vee R_{\!f}(\ldots, x_1, \ldots, x_n, \ldots, y_2, \ldots) \longleftarrow \nonumber\\
&R(\ldots, x_1, \ldots, x_n, \ldots, y_1, \ldots),  R(\ldots, x_1, \ldots, x_n, \ldots, y_2, \ldots), y_1 \neq y_2.
\end{eqnarray}
For FDs we do  not need the annotation $\mathbf{t}$, because inconsistencies are resolved by tuple deletions. For the same
reason we do not need program constraints.

\begin{example} \label{ex:first} (example \ref{ex:intro} continued)
 The
repair program $\Pi(D,\IC)$ is:
\begin{eqnarray}
&& P_{\!f}(x,y) \vee P_{\!f}(x,z)  \leftarrow P(x,y),
P(x,z), y \neq z. \label{eq:rep1} \\
&& P\ds(x,y)  \leftarrow  P(x,y), not~
P_{\!f}(x,y). \label{eq:rep2}\\
&& P(a,b). ~~P(a,c). ~~P(d,e). \label{eq:facts}
\end{eqnarray}
The first rule indicates that whenever there is a pair of tuples in conflict w.r.t. the FD, then
one of the tuples has to be deleted from the database. The second rule allows to collect the
tuples that remain in a repair after all conflicts have been solved after tuple deletions.
The two repairs can be obtained as the restrictions of the two stable models to their
$P\ds$ predicate: $D_1 = \{P\ds(a,b),
P\ds(d,e)\}$ and $D_2 =
\{P\ds(a,c), P\ds(d,e)\}$.  In the first one we have deleted the
tuple $P(a,c)$ from the database, and in the second one, the
tuple $P(a,b)$.

A possible  query is ${\cal Q}(x,y): P(x,y)$, which can be
represented by the query program $\Pi^\Q$:\begin{equation}
\label{eq:query} \A(x,y) \leftarrow P\ds(x,y).
\end{equation}
If $\Pi$ is the  program consisting of rules
(\ref{eq:rep1})-(\ref{eq:facts}), (\ref{eq:query}), the
consistent answers to query ${\cal Q}$ are those tuples
$\bar{a}$ of constants in $U$, such that $\Pi
\models_{\nit{cs}} \A(\bar{a})$.  In this case, the only consistent answer is $(d,e)$.\boxtheorem
\end{example}
We can see that repair programs for FDs  are stratified disjunctive
programs \cite{przy88}. They are also HCF programs, which makes it possible to translate them into equivalent normal (non-disjunctive)
programs \cite{BenDechter94,DEGV2001}. However, they are not stratified as normal programs.

\vspace{-3mm}
\paragraph{\bf \em Circumscription.} \ It was introduced in \cite{M1980} for the formalization of commonsense knowledge representation and reasoning via predicate (extension) minimization. Different forms of minimization can be accommodated. See \cite{L1994,KRHB} for  more recent presentations.

Let $\bar{P}, \bar{Q}$ be disjoint tuples of FO predicates. The circumscription of
$\bar{P}$ w.r.t. $\preceq$
in the FO sentence $\Sigma(\bar{P},\bar{Q})$ with variable $\bar{Q}$ can be expressed by means of the
SO sentence \cite{M1980,Lifschitz1985} $\nit{Circ}(\Sigma(\bar{P},\bar{Q});\bar{P};\bar{Q})$: \
$\Sigma(\bar{P},\bar{Q}) \wedge \neg \exists \bar{X} \bar{Y}(\Sigma(\bar{X},\bar{Y}) \wedge \bar{X} \preceq \bar{P} \wedge \bar{X} \neq \bar{P}),$
where $\bar{X}, \bar{Y}$ are tuples of SO variables that replace $\bar{P}$, resp. $\bar{Q}$ in $\Sigma(\bar{P},\bar{Q})$,
producing $\Sigma(\bar{X},\bar{Y})$.

Here, $\preceq$ stands for a FO definable pre-order relation (reflexive and transitive) between tuples of predicate extensions.
All the other predicates in $\Sigma(\bar{P},\bar{Q})$ are left untouched and they
are kept fixed during the minimization of those in $\bar{P}$, while those in $\bar{Q}$ become flexible.
By appropriately choosing the relation $\preceq$,  different forms of circumscription can be captured.  Prioritized circumscription is based on a prioritized partial order
relation between tuples $\bar{S} = (S_1, \ldots, S_m)$, and
$\bar{T} = (T_1, \ldots, T_m)$ of similar predicates (i.e. same length and
corresponding arities). It can be defined by $\bar{S} \preceq^\nit{pri} \bar{T} \ \equiv \ \bigwedge_{i=1}^m (\bigwedge_{j=1}^{i-1} S_i = T_i \rightarrow S_i \leq T_i)$.
Here, $\leq$ stands for the subset relation. The parallel circumscription of the predicates in $\bar{P}$ can be obtained by means of the relation: \
$\bar{S} \preceq^\nit{par} \bar{T} \ \equiv \ \bigwedge_{i=1}^m S_i \leq T_i$.

\vspace{-3mm}
\section{Second-Order Specification of Repairs}\label{sec:SO}

\vspace{-3mm}
In \cite{FLL2007,ExtFerraris}, the stable model semantics of
logic programs introduced in \cite{GL1991} is reobtained via
an explicit  specification in classical SO
predicate logic that is based on circumscription.

First, the program $\Pi$ is
transformed into (or seen as) a FO sentence $\psi(\Pi)$.
Next, the latter is
transformed into a SO sentence $\Phi(\Pi)$. Here,
$\psi(\Pi)$ is obtained from
$\Pi$ as follows: (a)
Replace every comma by $\wedge$, and every $\nit{not}$ by
    $\neg$.
(b) Turn every rule $\nit{Head} \leftarrow \nit{Body}$ into the
formula
    $\nit{Body}
    \to
    \nit{Head}$. (c) Form the conjunction of the universal closures
    of those formulas.

Now, given a FO sentence $\psi$ (e.g. the $\psi(\Pi)$ above), a SO sentence $\Phi$ is defined as
%\begin{equation}\label{eq:st}
$\psi \wedge \neg \exists \bar{X}((\bar{X}<
\bar{P}) \wedge \psi^\circ(\bar{X}))$, %\end{equation}
 where
$\bar{P}$ is the list of all  predicates
$P_1,...,P_n$ in $\psi$ that will be circumscribed, and $\bar{X}$ is a list of distinct
predicate variables $X^{\!P_1},...,X^{\!P_n}$, with $P_i$ and $X^{\!P_i}$ of the
same arity.
Here, $(\bar{X}< \bar{P})$ means
$(\bar{X}\leq \bar{P}) \wedge (\bar{X} \neq
\bar{P})$, i.e. $\bigwedge_i^n \forall \bar{x}(X^{\!P_i}(\bar{x})
\rightarrow P_i(\bar{x})) \wedge \bigvee_i^n(X^{\!P_i} \neq P_i)$. ~ $X^{\!P_i} \neq P_i$ stands for
$\exists \bar{x}_i(P_i(\bar{x}_i) \wedge \neg X^{\!P_i}(\bar{x}_i))$.

$\psi^\circ(\bar{X})$ is defined recursively as follows: (a)
$P_i(t_1,...,t_m)^\circ := X^{\!P_i}(t_1,...,t_m)$. (b)
    $(t_1=t_2)^\circ := (t_1=t_2)$. (c) $\perp ^\circ :=\perp$. (d) $(F
    \odot G)^\circ :=(F^\circ \odot G^\circ)$ for $\odot \in \{\wedge,
    \vee\}$. (e) $(F \rightarrow G)^\circ :=(F^\circ \rightarrow G^\circ)
    \wedge (F \to G)$. (f) $(QxF)^\circ := QxF^\circ$ for $Q \in
    \{\forall, \exists\}$. \ Notice that we assume there is no explicit logical negation in formulas. Instead, a formula of the form \ $\neg \chi$ \ is assumed to be represented as \ $(\chi \rightarrow \bot)$, with $\bot$ standing for an always false propositional formula.

     The Herbrand models of the SO sentence $\Phi(\Pi)$ associated to $\psi(\Pi)$
    correspond to the stable models of the original program $\Pi$ \cite{FLL2007}.\footnote{In \cite{FLL2007}, any FO sentence $\psi$ is
syntactically associated   to a SO sentence $\Phi$, and the stable models of $\psi$ are {\em defined} as the Herbrand models of $\Phi$.
If this process is applied to $\psi(\Pi)$, we reobtain the usual stable models of $\Pi$.} We can see that $\Phi(\Pi)$ is similar to a parallel circumscription of
    the predicates  in program $\Pi$ w.r.t. the FO
    sentence $\psi(\Pi)$ associated to $\Pi$ \cite{M1980,L1994}. In principle, the
    transformation rule (e) above could make formula $\Phi(\Pi)$ differ from a
    circumscription.

Now, let $D$ be a relational database, $\Pi^r$   the repair
program without the database facts, and  ${\cal Q}(\bar{x})$ a query represented by
a (stratified) non-recursive and normal Datalog$^\nit{not}$ query  $\Pi^\Q$ with answer predicate
$\nit{Ans}^\Q(\bar{x})$ (which appears only in heads of the program). From now on, \begin{equation}\Pi = D \cup \Pi^r
\cup \Pi^\Q \label{eq:pcqa}\end{equation} denotes the program
that can be used to obtain the consistent answers to $\Q$.
That is, $\Pi = \Pi(D,\IC) \cup \Pi^\Q$. Notice that $\Pi^r$ depends only on the ICs, and it includes
definitions for the annotation predicates. The only predicates
that can be shared by $\Pi^r$ and $\Pi^\Q$ are those of the
form $P\ds$, with $P \in {\cal S}$, and these appear only in
the bodies of the rules of $\Pi^\Q$. These predicates produce a {\em splitting} of the combined program, whose
stable models are obtained as extensions of the stable models for $\Pi(D,\IC)$  \cite{Splitting}.  In Example \ref{ex:first},
$\Pi^r$ is formed by rules (\ref{eq:rep1}) and (\ref{eq:rep2});
$D$ is the set of facts in (\ref{eq:facts}); and $\Pi^\Q$ is
(\ref{eq:query}).

The splitting of $\Pi$ mentioned in the previous paragraph allows us to analyze separately
$\Pi^r$ and $\Pi^\Q$. Since the latter is a non-recursive normal program, it is stratified, and its only stable
model (over a give extension for its extensional predicates) can be obtained by predicate completion, or
a prioritized circumscription \cite{przy88}.\footnote{The completion of a stratified program may have
models different from the standard model. Cf. \cite[pag. 139]{apt}, but those examples have recursion.}
Actually, if the query
is given directly as FO query, we can use instead
of the completion (or circumscription) of its associated program, the FO query itself.
In consequence, in the rest of this section we will concentrate mostly on the facts-free repair program $\Pi^r$.

In the following, we will usually omit the program constraints from
the repair programs, because their transformation via the SO
sentence of the program is straightforward: We obtain as
a conjunct of the SO sentence, the sentence $\forall \bar{x}
\neg (P_t(\bar{x}) \wedge P_{\!f}(\bar{x}))$
\cite[Prop. 2]{ExtFerraris} whenever  a program constraint of the form~
$\leftarrow P_t(\bar{x}), P_{\!f}(\bar{x})$ ~is required in the repair program.

\begin{example} (example \ref{ex:first} continued) \label{ex:2nd}
The first transformation step
of program $\Pi$ gives the FO formula $\psi(\Pi)$: \vspace{-5mm}
\begin{eqnarray}
P(a,b)
\wedge P(a,c) \wedge P(d,e) ~\wedge  \nonumber \\
\forall xyz( ((P(x,y)
\wedge P(x,z) \wedge y \neq z) \to (P\!\!_f(x,y) \vee
P\!\!_f(x,z))) ~\wedge \nonumber \\ \forall xy((P(x,y) \wedge \neg
P\!\!_f(x,y)) \to P\ds(x,y)) ~\wedge~  \forall xy( P\ds(x,y) \to
\nit{Ans}(x,y)). \label{eq:psi}
\end{eqnarray}
The second-order formula $\Phi(\Pi)$ that captures the
stable models of the original program is the conjunction of
(\ref{eq:psi}) and ~(with $<$ below being the ``parallel" pre-order \cite{Lifschitz1985}):

\begin{eqnarray}
&&\neg \exists
X^{\!P} X^{\!P}_f X^{\!P}\!\!\!\ds X^{\!\nit{Ans}}~[~(X^{\!P}, X^{\!P}_f, X^{\!P}\!\!\!\ds, X^{\!\nit{Ans}}) <
(P,P\!_f,P\ds,\nit{Ans}) ~\wedge \nonumber \ignore{\label{poto}}\\
&&X^{\!P}(a,b) \wedge X^{\!P}(a,c) \wedge X^{\!P}(d,e) ~\wedge \nonumber \ignore{\label{poto1}}\\
&&\forall xyz (X^{\!P}\!(x,y) \wedge X^{\!P}\!(x,z) \wedge y \neq z \to
X^{\!P}{\!\!\!_f}(x,y) \vee X^{\!P}{\!\!\!_f}(x,z)) ~~\wedge \nonumber \ignore{\label{poto2}}\\
&&\forall x y z(P(x,y) \wedge P(x,z) \wedge y \neq z \to P\!_f(x,y) \vee
P\!_f(x,z)) ~~\wedge \label{poto3}
\\&&\forall xy(X^{\!P}(x,y) \wedge (\neg P{\!\!_f}(x,y))^\circ \to X^{\!P}{\!\ds}(x,y)) ~~\wedge \label{eq:prime}\\
&&\forall xy(P(x,y) \wedge \neg P\!_f(x,y)) \to P\ds(x,y)) ~~\wedge \label{poto4} \\
&&\forall xy(X^{\!P}{\!\!\!\ds}(x,y) \to X^{\!\nit{Ans}}(x,y)) ~~\wedge \nonumber\\ &&\forall xy(P\ds(x,y)
\to \nit{Ans}(x,y) )]. \label{poto5}
\end{eqnarray}
In this sentence, the conjuncts (\ref{poto3}), (\ref{poto4}) and (\ref{poto5}), that already appear
in (\ref{eq:psi}), can be eliminated. The formula $(\neg P{\!\!_f}(x,y))^\circ$ in
 (\ref{eq:prime}) has to be expressed as $(P{\!\!_f}(x,y) \rightarrow \bot)^\circ$.
It turns out that, being the $\circ$-transformation of a negative formula,  it can be replaced by its original version without predicate variables,
i.e. by $\neg P{\!\!_f}(x,y)$
\cite[Prop. 2]{FLL2007}. We obtain that
$\Phi(\Pi)$ is logically equivalent to the conjunction of the UNA and DC sentences\footnote{From now on, unless stated otherwise, the UNA and DCA will be always
implicitly considered.} with (\ref{eq:psi}) and:
\begin{eqnarray}
&&\neg \exists
X^{\!P} X^{\!P}_f X^{\!P}\!\!\!\ds X^{\!\nit{Ans}}~[~(X^{\!P}, X^{\!P}_f, X^{\!P}\!\!\!\ds, X^{\!\nit{Ans}}) <
(P,P\!_f,P\ds,\nit{Ans}) ~\wedge\\
&&X^{\!P}(a,b) \wedge X^{\!P}(a,c) \wedge X^{\!P}(d,e) ~\wedge\\
&&\forall xyz (X^{\!P}\!(x,y) \wedge X^{\!P}\!(x,z) \wedge y \neq z \to
X^{\!P}{\!\!\!_f}(x,y) \vee X^{\!P}{\!\!\!_f}(x,z)) ~~\wedge\\
&&\forall xy(X^{\!P}(x,y) \wedge \neg P{\!\!_f}(x,y) \to X^{\!P}{\!\ds}(x,y)) ~~\wedge\\
&&\forall xy(X^{\!P}{\!\!\!\ds}(x,y) \to X^{\!\nit{Ans}}(x,y))].
\end{eqnarray}
Applying standard simplification techniques for second-order quantifiers \cite{Lifschitz1985,L1994}, $\Phi(\Pi)$ becomes
logically equivalent to
\begin{eqnarray}
&&\forall xy(P(x,y) \equiv (x=a \wedge y=b) \vee
(x=a \wedge y=c) \vee (x=d \wedge y=e)) ~\wedge \label{eq:e1}\\
&& \forall xy(P\ds(x,y) \equiv \nit{Ans}(x,y))  ~\wedge \label{eq:e2}\\
&&  \forall xy ((P(x,y) \wedge \neg P\!_f(x,y)) \equiv
P\ds(x,y)) ~\wedge \label{eq:e3} \\
&&   \forall xyz (P(x,y) \wedge P(x,z) \wedge y
\neq z \to (P\!_f(x,y) \vee P\!_f(x,z)))~~ \wedge \label{eq:e4}\\
&&\hspace*{-7mm}\neg \exists U\!_f((U\!_f < P\!_f) \wedge
\forall xyz(P(x,y) \wedge P(x,z) \wedge y \neq z \to
%\nonumber\\&&\phantom{a very very very very very very very}
(U\!_f(x,y) \vee U\!_f(x,z))). \label{eq:e5}
\end{eqnarray}
Here, $U\!_f < P\!_f$ stands for the formula $\forall
xy(U\!_f(x,y) \rightarrow P\!_f(x,y)) \wedge \exists x y
(P\!_f(x,y) \wedge \neg U\!_f(x,y))$. In this sentence, the
minimizations of the predicates $P, P\ds$ and
$\nit{Ans}$ are expressed as their predicate completion.
Predicate $P\!_f$ is minimized via (\ref{eq:e5}). \boxtheorem
\end{example}
In this example we have obtained the SO sentence for program
$\Pi$ as a parallel circumscription of the predicates in the repair program seen as a FO sentence. Even more, the circumscription
actually becomes a {\em prioritized circumscription} \cite{Lifschitz1985} given the stratified nature of the repair program: first the database predicate
is minimized, next $P\!_f$, next $P\ds$, and finally $\A$.

More precisely,  as we state in Proposition \ref{prop:strat},
repair programs, in their predicated-annotation version,
 become {\em stratified disjunctive Datalog programs}
 \cite{eiterGottlob97,przy88} in the absence of program denials\footnote{The latter spoil the stratification, because they have to be replaced by rules of
 the form $p \leftarrow P\!_t(\bar{x}), P\!_f(\bar{x}), \n p$.}. Since the latter, if needed, can be added at the end, after producing a
 circumscription or the SO stable sentence, we are left with a stratified disjunctive program.
\begin{proposition} \label{prop:strat} \em  For universal integrity constraints, repairs programs without their program constraints are stratified, and the
upwards stratification is as follows: 0. Extensional database predicates $P \in {\cal S}$; ~1. Predicates of the form
$P\!_f, P\!_t, P\s$; and ~2. Predicates of the form $P\ds$. \boxtheorem
\end{proposition}
Actually, this proposition can be extended, as proved in \cite{monicaJournal08} in its general form, to the case where $\IC$ includes an acyclic  set of
referential integrity constraints.
If a stratified query program is run on top of the repair program, the combined program becomes stratified, with the
stratification of the query on top of the one of the repair program. It is worth noticing that the data complexity of cautious query evaluation from
disjunctive logic programs with stratified negation is the same as for disjunctive
logic programs with unstratified negation and stable model semantics, namely $\Pi_2^P$-complete \cite{eiterGottlob97}

The stable models of the combined (stratified and disjunctive) program $\Pi$ coincide with the {\em perfect models} of the program
\cite{przy91}, and the latter can be
obtained as the (Herbrand) models of a prioritized circumscription that follows the stratification of the program \cite{przy88}. In consequence, we obtain the following
\begin{proposition} \label{prop:circ} \em
For a set of universal ICs, the SO sentence $\Phi$ associated to a repair program
$\Pi(\IC,D)$  is logically
equivalent to
\begin{eqnarray}
\R(D) &\wedge& \bigwedge_{P \in {\cal S}} \forall
\bar{x}((P(\bar{x}) \vee P_t(\bar{x})) \equiv
P\s(\bar{x})) \wedge \bigwedge_{P \in {\cal S}} \forall
\bar{x}(P\s(\bar{x}) \wedge \neg P_f(\bar{x}) \equiv
P\ds(\bar{x})) \nonumber\\
&\wedge& \bigwedge_{P \in {\cal S}} \forall \bar{x}
\neg (P_t(\bar{x}) \wedge P_{\!f}(\bar{x})) \wedge
\nit{Circ}(\Theta;\{P_t,P_{\!f}~|~P \in
\mathcal{S}\};\{P\s~|~P \in
\mathcal{S}\}). \label{eq:closure}
\end{eqnarray}
Here, the last conjunct is the {\em parallel circumscription} \cite{Lifschitz1985} of the
predicates in the second argument (with variable $P\s$ predicates) w.r.t. the theory $\Theta$
obtained from the conjunction rules in the repair program that are relevant to compute the $P\!_t, P\!_f$'s,  seen as FO sentences.\footnote{They are rules 1.- 3. in Example
\ref{ex:genProg}.}
 \boxtheorem
\end{proposition}
This result has been obtained from the stratification of the repair programs. However, it is possible to obtain the same
result by simplifying the SO sentence associated to it as done in Example \ref{ex:2nd}. Notice that the more involved repair
program in Example \ref{ex:genProg} already contains the relevant features of a general repair program for universal ICs, namely
the negations in the rule bodies affect only base predicates and the predicates $P\s$ in the definitions of the $P\ds$ \cite{twelve}.

In any case, we obtain a SO specification of the logic program for CQA $\Pi$  in (\ref{eq:pcqa}), which combined with
(\ref{eq:cqaLP}), gives
\begin{equation}D
\models_c \Q(\bar{a}) ~~\Longleftrightarrow~~
\Phi(\Pi) \models \A^{\!\Q}(\bar{a}),
\end{equation}
where $\Phi(\Pi)$ is the SO sentence which captures the stable models
of $\Pi$.\footnote{If we omit the DCA and UNA axioms,
on the RHS the logical consequence
is relative to Herbrand models.} Actually, $\Phi(\Pi)$ can be decomposed as the conjunction of three formulas:
\begin{proposition} \label{prop:charCQA} \em
Let $\Phi$ be the SO sentence for the program $\Pi$ in
(\ref{eq:pcqa}) for CQA. It holds:~

\vspace{2mm}
\centerline{$D \models_c
\Q(\bar{a}) ~\Longleftrightarrow~ \{\R(D), \Phi(\Pi^r),
\Phi(\Pi^\Q)\} \models \A(\bar{a}).$}

 \vspace{2mm} \noindent
Here, $\Phi(\Pi^r)$ is a SO sentence that specifies the repairs for fixed extensional predicates, and  $\Phi(\Pi^\Q)\}$
a SO sentence that specifies the models of the query, in particular predicate $\A^\Q$, for fixed predicates $P\ds$.
\boxtheorem
\end{proposition}
\begin{example} (example \ref{ex:2nd} continued)~ $\R(D)$ is captured by the DCA, UNA plus (\ref{eq:e1});
$\Phi(\Pi^r)$ by (\ref{eq:e3})-(\ref{eq:e5}); and $\Phi(\Pi^\Q)$ by
(\ref{eq:e2}). Actually, what we have obtained is that for consistent answers $(t_1,t_2)$, it holds
\begin{equation} \label{eq:SOquery}
\Psi \wedge \forall x \forall y(\A(x,y) \equiv P\ds(x,y)) ~\models~ \A(t_1,t_2),
\end{equation}
where $\Psi$ is the SO sentence that is the conjunction of (\ref{eq:e1}), (\ref{eq:e3})-(\ref{eq:e5}). \boxtheorem
\end{example}
We can see that we have transformed the problem of CQA into a problem of reasoning in classical SO predicate logic.
Most commonly the query $\Q$ will be given as a FO query or as a safe and non-recursive Datalog$^\nit{not}$
program. In these cases, $\Phi(\Pi^\Q)$ is obtained by predicate completion and will contain as a conjunct an explicit definition of predicate
$\A^\Q$. The definition of $\A^\Q$ will be of the form  $\forall \bar{x} (\A^\Q(\bar{x}) \equiv \Psi(\bar{x}))$, where $\Psi(\bar{x})$ is a FO formula
containing only predicates of the form $P\ds$, with $P \in S$, plus possibly some built-ins and auxiliary predicates.
For example, in (\ref{eq:SOquery}) we have an explicit definition of $\A$. As another example,
for the FO query ${\cal Q}(x)\!:
P(x) \wedge \neg \exists y R(x,y)$, the query program has
two rules: $\A^\Q(x) \leftarrow P\ds(x), \n B(x)$, and $B(x)
\leftarrow R\ds(x,y)$, with an auxiliary predicate $B$. $\Phi(\Pi^\Q)$ is the conjunction of
$\forall x(\A^\Q(x) ~\equiv~
P\ds(x) \wedge \neg B(x))$ and $\forall x(B(x) ~\equiv~ \exists y
R\ds(x,y))$.

\vspace{-2mm}
\section{From Second-Order to First-Order CQA under FDs} \label{sec:QE}

\vspace{-2mm}
%\comlb{Relate to Beth's definability theorem. It does not hold for finite models. What about here? See Jouko Vaananen's slides.}

We discuss in this section the possibility of using a program
for CQA $\Pi$ of the form (\ref{eq:pcqa}) to obtain  a FO theory from which to do CQA as classical entailment. In particular,
exploring  the possibility of obtaining a FO rewriting of the original query.
 The idea is to do it through the analysis of the SO sentence associated to the program.
In order to explore the potentials of this approach, we restrict ourselves to the case of FDs, the
most studied case in the literature w.r.t. complexity of CQA \cite{CM2005,FM2007,wijsenSurvey}.

We start with a schema with a predicate $P(X,Y)$, with the $\nit{FD}: ~X \rightarrow Y$, as in Example
\ref{ex:intro}. The repair program $\Pi(D,\nit{FD})$
of an instance $D$ w.r.t. $\nit{FD}$ is associated to
 the
circumscription of $\p$ given by the conjunction of (\ref{eq:e1}), (\ref{eq:e3})-(\ref{eq:e5}).
We concentrate on the last conjunct, (\ref{eq:e5}), which can be
expressed as
  \begin{equation}\label{eq:myCirc} \neg \exists U\!_f((U\!_f <
P\!_f) \wedge \forall xyz(\kappa(x,y,z) \to (U\!_f(x,y) \vee
U\!_f(x,z))),
\end{equation}
where $\kappa(x,y,z)$ is the formula $P(x,y) \wedge P(x,z)
\wedge y \neq z$, that captures the inconsistencies w.r.t. \F.

We will apply to (\ref{eq:myCirc}) the techniques for elimination of SO quantifiers
 developed in \cite{DLS1997} on the basis of Ackerman's
Lemma \cite{Ack35,Ack54}.
First of all, we express (\ref{eq:myCirc}) as an equivalent  universally
quantified formula (for simplicity, we use $U$ instead of $U_{\!f}$):
\begin{equation}\label{eq:myCirc2}
\forall U (\forall x y z(\kappa(x,y,z) \rightarrow U(x,y) \vee
U(x,z)) \wedge U \leq \p ~\rightarrow~ \p \leq U).
\end{equation}
Its negation produces the existentially quantified formula
\begin{equation}\label{eq:myCirc3}
\exists U (\forall x y z(\kappa(x,y,z) \rightarrow U(x,y) \vee
U(x,z)) \wedge U \leq \p \wedge \neg \p \leq U).
\end{equation}
We obtain the following sequence of logically equivalent
formulas
\begin{eqnarray}
\exists U (&&\forall x y z(\neg \kappa(x,y,z) \vee U(x,y) \vee
U(x,z)) \wedge \forall uv(\neg U(u,v) \vee \p(u,v)) \nonumber \\
&&\phantom{espacio muuuuuuuuuuy largo}\wedge \exists st(\p(s,t) \wedge \neg U(s,t))). \nonumber \\
\exists st\exists U(&&\forall xyz(\neg\kappa(x,y,z) \vee U(x,y)
\vee U(x,z)) \wedge \label{eq:a}\\&& \phantom{espacio largo}\forall uv(\neg U(u,v)
\vee \p(u,v)) \wedge (\p(s,t) \wedge \neg U(s,t))). \nonumber
\end{eqnarray}
The first conjunct in (\ref{eq:a}), with $w =\vee(y,z)$
standing for $(w = y \vee w = z)$, can be equivalently written as any of the following
(also equivalent) formulas \\
\\
$\forall xyz(\neg \kappa(x,y,z) \vee \exists w(w = \vee(y,z)
\wedge U(x,w))).$\\
$\forall xyz\exists w(\neg \kappa(x,y,z) \vee (w = \vee(y,z)
\wedge U(x,w))).$\\
$\forall xyz\exists w((\neg \kappa(x,y,z) \vee w = \vee(y,z))
\wedge (\neg \kappa(x,y,z) \vee U(x,w))).$\\
$\forall xyz\exists w((\neg \kappa(x,y,z) \vee w = \vee(y,z))
\wedge \forall r(\neg \kappa(x,y,z) \vee r \neq w \vee U(x,r))).$\\
$\exists f\forall xyz\forall r((\neg \kappa(x,y,z) \vee f(x,y,z)
= \vee(y,z))
\wedge$\\ \hspace*{6cm}$(\neg \kappa(x,y,z) \vee r \neq f(x,y,z) \vee U(x,r))).$\\
$\exists f\forall r(\forall xyz(\neg \kappa(x,y,z) \vee f(x,y,z)
= \vee(y,z))
\wedge$\\\hspace*{6cm} $\forall xyz(\neg \kappa(x,y,z) \vee r \neq f(x,y,z) \vee U(x,r))).$\\
$\exists f\forall r(\forall x_1y_1z_1(\neg \kappa(x_1,y_1,z_1)
\vee f(x_1,y_1,z_1) = \vee(y_1,z_1))
\wedge$\\\hspace*{6cm} $\forall xyz(\neg \kappa(x,y,z) \vee r \neq f(x,y,z) \vee U(x,r))).$\\
\\
Above, $\exists f$ is a quantification over functions. Formula (\ref{eq:a}) becomes\\
\\
$\exists st\exists U(\exists f \forall r (\forall x_1y_1z_1(\neg
\kappa(x_1,y_1,z_1) \vee f(x_1,y_1,z_1) = \vee(y_1,z_1))
\wedge$\\ \hspace*{3cm} $\forall xyz(\neg \kappa(x,y,z) \vee r \neq f(x,y,z)
\vee U(x,r))) \wedge$\\ \hspace*{4cm} $\forall uv(\neg U(u,v) \vee \p(u,v))
\wedge (\p(s,t) \wedge \neg U(s,t))).$\\
Equivalently, ~$\exists st\exists f \exists U\forall x \forall r((\forall
x_1y_1z_1(\neg \kappa(x_1,y_1,z_1) \vee f(x_1,y_1,z_1) =
\vee(y_1,z_1)) \wedge$\\ \hspace*{3cm}$\forall yz(\neg
\kappa(x,y,z) \vee r \neq f(x,y,z) \vee U(x,r))) \wedge$\\
\hspace*{4cm}$\forall uv(\neg U(u,v) \vee \p(u,v)) \wedge (\p(s,t) \wedge
\neg U(s,t))).$

\noindent Now we
are ready to apply Ackermann's lemma. The last formula can be
written as
\begin{equation} \label{eq:ready}
\exists st\exists f \exists U\forall x \forall r ((A(x,r) \vee
U(x,r)) \wedge B(\neg U \mapsto U)).
\end{equation}
Here, $B(\neg U \mapsto U)$ indicates the formula $B$ where
predicate $U$ has been replaced by $\neg U$. Formulas $A, B$
are as follows\\
\\
$A(x,r):~~ \forall yz(\forall yz(\neg \kappa(x,y,z) \vee r \neq
f(x,y,z)).$\\
$B(U):~~~~ \forall x_1y_1z_1(\neg \kappa(x_1,y_1,z_1) \vee
f(x_1,y_1,z_1) = \vee(y_1,z_1)) ~\wedge$\\ \hspace*{5cm} $\forall uv(U(u,v)
\vee \p(u,v)) \wedge (\p(s,t) \wedge U(s,t))).$

\noindent Formula $B$ is positive in $U$, in consequence, the whole
subformula in (\ref{eq:ready}) starting with $\exists U$ can be
equivalently replaced by $B(A(x,r) \mapsto U)$ \cite[lemma 1]{DLS1997}, getting rid of
the SO variable $U$, and thus
obtaining\\
\\
$\exists st\exists f(\forall x_1y_1z_1(\neg \kappa(x_1,y_1,z_1)
\vee f(x_1,y_1,z_1) = \vee(y_1,z_1)) \wedge$\\ \hspace*{3cm} $\forall
uv(\forall yz(\neg\kappa(u,y,z) \vee v \neq f(u,y,z) \vee
\p(u,v)) \wedge$\\ \hspace*{5cm} $(\p(s,t) \wedge \forall y_1 z_1(\neg
\kappa(s,y_1,z_1) \vee t \neq f(s,y_1,z_1))).$\\
Equivalently, ~~$\exists st\exists f(\forall x_1y_1z_1(\neg \kappa(x_1,y_1,z_1)
\vee f(x_1,y_1,z_1) = \vee(y_1,z_1)) \wedge$\\ \hspace*{3cm}$\forall
uvyz(\neg\kappa(u,y,z) \vee v \neq f(u,y,z) \vee \p(u,v))
\wedge$\\ \hspace*{4cm}$(\p(s,t) \wedge \forall y_1 z_1(\neg \kappa(s,y_1,z_1)
\vee t \neq f(s,y_1,z_1))).$\\
Equivalently, ~$\exists st\exists f\forall xyz((\neg \kappa(x,y,z) \vee
f(x,y,z) = \vee(y,z)) \wedge$\\ \hspace*{3cm}$(\neg\kappa(x,y,z) \vee
\p(u,f(x,y,z))) \wedge$\\ \hspace*{4cm}$ (\p(s,t) \wedge ( x \neq s \vee \neg
\kappa(x,y,z) \vee t \neq f(x,y,z)))).$

\noindent Now we unskolemize, getting rid of the function variable $f$, obtaining\\
$\exists st\forall xyz \exists w((\neg \kappa(x,y,z) \vee w =
\vee(y,z)) \wedge (\neg\kappa(x,y,z) \vee \p(u,w)) \wedge$\\ \hspace*{5cm}
$(\p(s,t) \wedge ( x \neq s \vee \neg \kappa(x,y,z) \vee t \neq
w))).$

\noindent This formula is logically equivalent to the negation of
(\ref{eq:myCirc2}). Negating again, we obtain a formula that is
logically equivalent to (\ref{eq:myCirc2}), namely\\
$\forall s t \exists x y z \forall w((\kappa(x,y,z) \wedge w \neq y \wedge w \neq z) \vee
(\kappa(x,y,z) \wedge \neg \p(x,w)) \vee$\\\hspace*{5cm}$ (\neg \p(s,t) \vee (x =s \wedge \kappa(x,y,z)
\wedge t =w))$.\\
Equivalently, ~$\forall s t \exists x y z \forall w((\kappa(x,y,z) \wedge w \neq y \wedge w \neq z) \vee$\\
\hspace*{2cm}$(\kappa(x,y,z) \wedge \neg \p(x,w)) \vee \neg \p(s,t) \vee (x =s \wedge \kappa(x,y,z)
\wedge t =w))$.\\
Or,~
$\forall s t (\p(s,t) \rightarrow \exists x y z (\kappa(x,y,z) \wedge \forall w[(w \neq y \wedge w \neq z) \vee$\\
\hspace*{6cm} $\neg \p(x,w) \vee  (x =s \wedge t =w)])$.

\noindent The formula in the square bracket inside can be equivalently replaced by\\
\centerline{$((w =y \vee w = z) \wedge \p(x,w)) \rightarrow (s = x \wedge t = w)$.}\\
So, we obtain~ $\forall s t (\p(s,t) \rightarrow \exists x y z (\kappa(x,y,z) \wedge (\p(x,y) \rightarrow s=x \wedge t =y)
~\wedge$\\ \hspace*{6.9cm} $(\p(x,z) \rightarrow s=x \wedge t =z)))$.

Due to the definition of $\kappa(x,y,z)$, it must hold $y \neq z$. In consequence, we obtain\\
$\forall s t (\p(s,t) \rightarrow \exists z(\kappa(s,t,z) \wedge \neg \p(s,z)))$.
\begin{proposition} \label{prop:fofds} \em
Let $\IC$ be the FD
%\begin{equation}\label{eq:fds}
~$\forall x y z(P(x, y) \wedge
P(x,z) \rightarrow y = z)$.
%\end{equation}
The SO sentence for the repair program $\Pi(D,\IC)$ is
logically  equivalent to a FO sentence, namely to the
conjunction of
(\ref{eq:e1}), (\ref{eq:e3}) (i.e. the completions of the predicates $P,  P\ds$,  resp.), (\ref{eq:e4}),
and
\begin{equation}\label{eq:elim}
\forall s t (\p(s,t) \rightarrow \exists z(\kappa(s,t,z) \wedge \neg \p(s,z))),
\end{equation}
where $\kappa(x,y,z)$ is the formula that captures a violation of the
FD, i.e. $(P(x, y) \wedge
P(x, z) \wedge y \neq z)$.\boxtheorem
\end{proposition}
This is saying, in particular, that whenever there is a conflict between two tuples, one of them must be deleted, and for every deleted
tuple due to a violation, there must be a tuple with the same key value that has not been deleted. Thus, not all mutually conflicting tuples
can be deleted.

Now, reconsidering CQA,
 if we have a query $\cal Q$, we can obtain the consistent answers $\bar{t}$ as entailments in classical predicate logic
\begin{equation}\label{eq:FOqueryGen}
\psi \wedge \forall \bar{x} (\A^{\! \cal Q}\!(\bar{x})) \equiv \chi(\bar{x})) ~\models~ \A^{\!\cal Q}\!(\bar{t}),
\end{equation}
where $\psi$ is the FO sentence that is the conjunction of (\ref{eq:e1}), (\ref{eq:e3}), (\ref{eq:e4})
and (\ref{eq:elim}); and $\chi$ is the FO definition of $\A^{\cal Q}$ in terms of the $P\ds$ predicate. For example,
for
the query ${\cal Q}:~ P(x,y)$, we have, instead
of (\ref{eq:SOquery}):
\begin{equation}\label{eq:FOquery}
\psi \wedge \forall x \forall y(\A(x,y) \equiv P\ds(x,y)) ~\models~ \A(t_1,t_2).
\end{equation}
From here we obtain, using (\ref{eq:e3}), that $(t_1,t_2)$ is a  consistent answer
iff $\psi \models P\ds(t_1, t_2)$ iff  $\psi \models (P(t_1, t_2) \wedge \neg P_f(t_1,t_2))$.
That is,
\begin{eqnarray}\label{cqaFD}
&&\hspace*{-4mm}\{\R(D), \forall xyz (\kappa(x,y,z) \to (P\!_f(x,y) \vee P\!_f(x,z))), \nonumber
\\  &&~\forall x y (\p(x,y) \rightarrow \exists z(\kappa(x,y,z) \wedge \neg \p(x,z)))\} \models P(t_1, t_2) \wedge \neg P_f(t_1,t_2).
\end{eqnarray}
This requires $P(t_1,t_2)$ to hold in $\R(D)$, and the negation of $\neg P_f(t_1,t_2)$ to be inconsistent with the
theory on the LHS of (\ref{cqaFD}). This happens iff $\forall z \neg \kappa(t_1,t_2,z)$ follows from $\R(D)$. In consequence,  $(t_1,t_2)$ is
a consistent answer iff $\R(D) \models P(t_1,t_2) \wedge \forall z \neg \kappa(t_1,t_2,z)$, which is equivalent to
\begin{equation}\label{eq:rewr}
D \models P(t_1,t_2)
\wedge \neg \exists z (P(t_1,z) \wedge z \neq t_2).
\end{equation}
The rewriting in (\ref{eq:rewr}),  already presented in Example \ref{ex:intro}, is one of those
obtained in \cite{ABC1999} using a more general rewriting methodology for queries that are
quantifier-free conjunctions of database literals and classes of ICs that include FDs. The technique in
\cite{ABC1999} is not based on explicit specification of repairs. Actually, it relies on an iteration of resolution steps between ICs and
intermediate queries, and is not defined for queries or ICs with existential quantifiers.  Rewriting (\ref{eq:rewr}) is also a particular case
of a result in \cite[theo. 3.2]{CM2005} on FO rewritability of CQA for conjunctive queries without free variables.\footnote{That result can be applied
with our query ${\cal Q}(x,y): P(x,y)$, by transforming it  first into $\exists x \exists y(P(x,y) \wedge x=t_1 \wedge y = t_2)$, with generic, symbolic constants
$t_1,t_2$, as above.}

Notice that (\ref{eq:FOqueryGen}), in spite of being expressed as entailment in FO logic,  does not
necessarily allow us to obtain a FO rewriting to consistently answering query ${\cal Q}(\bar{x})$. A FO rewriting,
and the subsequent polynomial-time data complexity, are guaranteed when we obtain a condition of the form $D \models \varphi(\bar{t})$
for consistent answers $\bar{t}$, and $\varphi$ is a FO formula expressed in terms of the database predicates in $D$. This is different
from we could naively obtain from  (\ref{eq:FOqueryGen}), namely
a sentence containing possibly complex and implicit view definitions, like the derived definition of
$P_{\!f}$ above. A finer analysis from (\ref{eq:FOqueryGen}) is required in order to obtain a FO rewriting, whenever possible.

The particular case considered in Proposition \ref{prop:fofds} has all the features of the case of FDs most studied in the literature, namely where
there is one FD per database predicate \cite{CM2005,FM2007,wijsenSurvey}. Under this assumption, if we have a class of FDs involving different predicates, we can
treat each of the FDs separately, because there is no
interaction between them. So, each predicate
$P_f$ can be circumscribed independently from the
others, obtaining results similar to those for the particular case.

\vspace{-3mm}
\section{Towards Fixed-Point Logic}\label{sec:todo}

\vspace{-3mm}
As described in Section \ref{se:introduction}, there are syntactic classes of CQs for which consistent query answering can be done in polynomial time in data complexity. For one class, this can be done via FO query rewriting. For a different class,  its queries provably do not admit a first-order rewriting. Even more, one can decide if a CQ falls in this case or not \cite{wijsenSurvey,wijsenSigRec}.

For example, the Boolean conjunctive query $\mathcal{Q}\!: \ \exists x \exists y(R(x, y) \wedge S(y, x))$, with the first attributes of $R$ and $S$ as keys for them, is  a query in the second class in that it can be consistently answered in polynomial time, but no FO rewriting for it exists. Results of this kind are established in  \cite{wijsenIS09,wijsenIPL10} by means of the notions of {\em Hanf-locality} and {\em Ehrenfeucht-Fra\"iss\'e games} for FO-logic \cite{libkin}.

This opens the ground for investigating two problems:
\begin{enumerate}

\item Apply the second-order quantifier elimination technique in \cite{DLS1997}, that we applied in this work, with the purpose of recovering the FO rewritings for the whole class of queries that admit FO consistent rewritings (as determined by Wijsen \cite{wijsenIS09}).
\item Identify and obtain logical languages that can be used for rewriting the queries in the second class, in such a way that query answering for the rewritten query can be done in polynomial time.
\end{enumerate}
For the second problem, it would be interesting to see if second-order quantifier elimination could be applied to second-order specification of Section \ref{sec:SO}, in such a way that the resulting query is expressed, not in FO logic, but in fixed-point logic, which would lead to a polynomial-time answer \cite{libkin}.
\ Actually, in \cite{FI96}, the authors have been able to eliminate second-order quantifiers, obtaining  fixed-point formulas. It is worth investigating  if this is a way to obtain polynomial-time, logical,  but non-FO, rewritings for CQA. This undertaking is not a priori impossible. The existence of non-FO rewritable but PTIME-complete queries (in data) already identified \cite{wijsenIS09,Paraschos20} is in principle compatible with the PTIME-completeness of fixed-point logic (in data) \cite{DEGV2001}.

\vspace{-3mm}
\section{Conclusions} \label{sec:concl}
\vspace{-3mm}
Repair programs for consistent query answering have been well studied in the literature. They specify the database repairs as their stable models.
On their basis, and using
available implementations for the disjunctive stable model semantics for logic programs,\footnote{In \cite{monicaJournal08}, its is shown how to
use {\em DLV} \cite{leone06} for CQA.} we have the most general
mechanism for CQA \cite{monicaJournal08}. As expected, given the nature of CQA, its semantics is non-monotonic, and its logic is non-classical.
In this work we have presented the first steps of an ongoing research program that aims to take advantage of specifications of database repairs in
classical logic, from which CQA can be done as logical entailment.

The fact that stable models, and in particular database repairs, can be specified in SO logic can be obtained from complexity-theoretic results. The
decision problem of stable
model checking (SMC) consists in deciding if, for a fixed program, a certain finite input set of atoms is a stable model of the program. The
repair checking problem (RC) consists, for a fixed set of ICs $\IC$, if $D'$ if a repair of $D$ w.r.t. to \IC. Here, $D,D'$ are inputs to the
problem.  Both SMC and RC are $\nit{coNP}$-complete (cf. \cite{eiterGottlob97} and \cite{CM2005}, resp.). Since by Fagin's theorem (cf. \cite{fagin74} and \cite[chapter 9]{libkin}),
universal SO logic captures the class $\nit{coNP}$, there is a a universal SO sentence that specifies the repairs. For the same reason, the stable
models of a fixed program can be specified in universal SO classical logic. (Cf. also \cite{eiter97} for applications of such representation results.)

In this work we have shown concrete specifications of repairs in SO classical logic. They have been obtained from the results in \cite{FLL2007,ExtFerraris},
that presents a  characterization of the stable models as models of a theory in SO predicate logic. However,
due to the nature of repair programs, we are able to provide a circumscriptive SO characterization of them. A first and preliminary
circumscriptive approach to the specification of database repair was presented in \cite{amai04}.\footnote{The use of the circumscriptive, SO version of the stable models semantics has been also successfully applied to introduce non-stratified negation in Datalog$^\pm$ ontological languages \cite{pieris}.}

Furthermore, we have shown, starting from
the SO specification of stable models in \cite{FLL2007},  that, in the case of
repair programs w.r.t. functional dependencies, it is possible to obtain a specification in first-order classical logic. The FO theory can be obtained
from the circumscriptive theory by newer quantifier elimination methods that have their origin in the work of Herbrand on
decidable classes for the decision problem.   In particular, we have shown that it is possible to obtain first-order
first-order rewritings for CQA of the kind presented in \cite{ABC1999}.

Many problems are open for ongoing and future research. For example, and most prominently,
the natural question is as to whether the combination of a
repair program and a query program can be used, through their
transformation, to obtain more efficient algorithms that the
standard way of evaluating disjunctive
logic programs under the stable model semantics. We know that in the worst cases of CQA this is not
possible, but it should be possible for easier classes of queries and ICs.

More specifically, the following are natural problems to consider: (a) Identification of classes of ICs and queries
for which repair programs can be automatically ``simplified'' into queries of lower complexity.
In particular, reobtain previously identified classes, and identify new ones. (b)
More generally, obtain new complexity results for CQA. (c) Shed more light on those cases, possibly classes, where
CQA can be done in polynomial time, but not via FO rewriting.

Furthermore, the ``logic" of CQA is not fully understood yet. We should be able to
better understand the logic of CQA through the analysis of repair programs. However, their version in classical logic as
 presented in this work seems more appropriate for this task.
For example, we would like to
obtain results about compositionality of CQA, i.e. determining consisting answers to queries on the bases of
  consistent answers to subqueries or auxiliary views. Techniques of this kind are important for the practice of CQA.
  We know how to logically manipulate and transform a specification written in classical FO or SO logic, which is not
  necessarily the case for logic programs. It seems to be easier to (meta)reason about the specification if it is written in classical
logical than written as a logic program, which is mainly designed to compute from it.

  Also dynamic aspects of CQA have been largely neglected (cf. \cite{LoBe07} for some initial results). Computational
  complexity results and incremental algorithms for CQA are still missing. Results on updates of logic programs and/or theories
  in classical logic might be used in this direction.\\

  \noindent {\bf Acknowledgements:} \ Useful comments from anonymous reviewers for a previous version of this paper are much appreciated. Leopoldo Bertossi has been partially funded by the ANID - Millennium Science Initiative Program - Code
ICN17-002.

\appendix

\section{General Repair-Programs} \label{sec:app1}
For a
set $\IC$ of universal constraints of the form:~~
\begin{equation}\label{eq:format}
  \forall \bar{x}(\bigwedge_{i = 1}^{m} P_i(\bar{x}_i) ~\rightarrow~
\bigvee_{j=1}^n Q_{j}(\bar{y}_j) \vee
    \varphi),
\end{equation}
the repair program $\Pi(\IC,D)$ for a database instance $D$ without nulls  has the following rules:

\noindent 1. ~ Program facts: $P(\bar{a})$ for each atom
    $P(\bar{a}) \in D$.\\
    2. ~ For a constraint of the form (\ref{eq:format}), the rules:
\begin{equation*} \label{eq:preR2}
\begin{split}
\bigvee_{i=1}^n P_{i\mathbf{f}}(\bar{x}_i) \vee &
\bigvee_{j=1}^m Q_{j\mathbf{t}}(\bar{y}_j) \leftarrow
\bigwedge_{i=1}^n
P_{i\mathbf{\star}}(\bar{x}_i), \\
& \hspace*{2.4cm}\bigwedge_{Q_j \in Q'}
Q_{j\mathbf{f}}(\bar{y}_j), \bigwedge_{Q_k \in Q''} \nit{not}~
Q_k(\bar{y}_k), ~\bar{\varphi}.
\end{split}
\end{equation*}
This for every pair of sets $Q'$ and $Q''$ such that $Q' \cup
Q'' = \bigcup_{i=1}^m \{Q_i\}$, and $Q' \cap Q'' = \emptyset$.
Here $\bar{x}$ is the tuple of all variables appearing in
database atoms in the tuple, and $\bar{\varphi}$ is a
conjunction of
built-ins equivalent to the negation of $\varphi$.\\
3. ~ For each predicate $P \in \mathcal{S}$ the annotation
    rules:

\centerline{$P\s(\bar{x}) \leftarrow P(\bar{x})$ ~~and~~
$P\s(\bar{x}) \leftarrow P_\mathbf{t}(\bar{x})$.}

\noindent 4. ~ For every
predicate $P \in \mathcal{R}$, the interpretation
    rule:~
$P\ds(\bar{x}) \leftarrow P\s(\bar{x}), \nit{not}\
P_\mathbf{f}(\bar{x})$.\\
5. For each database predicate $P$, the program constraint:~
$\leftarrow P_\mathbf{t}(\bar{x}), P_\mathbf{f}(\bar{x})$.


\vspace{-3mm}
\begin{thebibliography}{10}
\bibitem{AHV1995} Abiteboul, S., Hull, R. and Vianu, V. \newblock {\em Foundations of Databases}. Addison-Wesley,
1995.

\bibitem{Ack35}
Ackermann, W. \newblock Untersuchungen \"uber das Eliminationsproblem der mathematischen Logik.
{\em Mathematische Annalen}, 1935, 110:390-413.

\bibitem{Ack54}
Ackermann, W. \newblock
{\em Solvable cases of the Decision Problem}.
North-Holland Pub. Co., 1954.

\bibitem{pieris}
Alviano, M.,
Morak, M. and
Pieris, A. \ Stable Model Semantics for Tuple-Generating
Dependencies Revisited. Proc. PODS, 2017, pp. 377-388.


\bibitem{apt}
Apt, K., Blair, H. and  Walker, A. \newblock Towards a Theory of Declarative Knowledge.
In {\em Foundations of Deductive Databases and Logic Programming}, J. Minker (ed.), Morgan Kaufmman, 1988,
pp. 89-148.

\bibitem{ABC1999} Arenas, M., Bertossi, L. and Chomicki, J. Consistent
Query Answers in Inconsistent Databases. {\em Proc. ACM Symposium
on Principles of Database Systems}, ACM Press, 1999, pp. 68-79.

\bibitem{ABC2003} Arenas, M., Bertossi, L. and Chomicki, J. Answer
Sets for Consistent Query Answering in Inconsistent Databases.
{\em Theory and Practice of Logic Programming}, 2003, 3(4-5):393-424.

\bibitem{BB2003} Barcelo, P. and Bertossi, L. \newblock Logic Programs for Querying
Inconsistent Databases. {\em Proc. Practical Aspects of
Declarative Languages}, Springer LNCS 2562, 2003, pp.
208-222.

\bibitem{BBB2003} Barcelo, P., Bertossi, L. and Bravo, L. \newblock Characterizing
and Computing Semantically Correct Answers from Databases with
Annotated Logic and Answer Sets. In {\em Semantics of Databases},
Springer LNCS 2582, 2003, pp. 1-27.


\bibitem{BenDechter94}
Ben-Eliyahu, R. and Dechter, R. \newblock Propositional Semantics for Disjunctive Logic Programs.
{\em Annals of Mathematics
and Artificial Intelligence}, 1994, 12(1-2):53-87.

\bibitem{amai04}
Bertossi, L. and Schwind, C.
\newblock Database Repairs and Analytic Tableaux. {\em Annals of Mathematics and Artificial Intelligence}, 2004, 40(1-2):5-35.

\bibitem{B2006} Bertossi, L. Consistent Query Answering in Databases. In {\em ACM
Sigmod Record}, June 2006, 35(2):68-76.

\bibitem{arequipa}
Bertossi, L.
From Database Repair Programs to Consistent Query Answering in Classical Logic (extended abstract). Proc. Alberto Mendelzon International Workshop on Foundations of Data Managemente (AMW),  2009. CEUR Workshop Proceedings, Vol. 450, 2009.

\bibitem{bertossi2011}
Bertossi, L. \
{\em Database Repairing and Consistent Query Answering}. \ Synthesis Lectures on Data Management, Morgan \& Claypool Publishers, 2011.

\bibitem{BB2006} Bravo, L., Bertossi, L. \newblock Semantically Correct
    Query Answers in the Presence of Null Values.
    {\em Proc. EDBT WS on Inconsistency and Incompleteness in
    Databases}, Springer LNCS 4254, 2006, pp. 336-357.

    \bibitem{brewka}Brewka, G., Eiter, T. and Truszczynski, M. \ Answer Set Programming at a Glance. \ {\em Communications of the ACM}, 2011, 54(12):92-103.

%\bibitem{darwicheEcai20}

%\bibitem{CEG1992}
%Cadoli, M., Eiter, T. and Gottlob, G. An Efficient Method for
%Eliminating Varying Predicates from
 %              a Circumscription. {\em Artificial Intelligence},
%1992, 54(3):397-410.

%\bibitem{CLR2003} Cali, A., Lembo, D. and Rosati, R. \newblock On the
%Decidability and Complexity of Query Answering over Inconsistent
%and Incomplete Databases. {\em Proc. ACM Symposium on
%Principles of Database Systems (PODS)}, ACM Press, 2003, pp.
%260-271.

\ignore{
\bibitem{CB2005} Caniupan, M., Bertossi, L. \newblock Optimizing Repair Programs for
Consistent Query Answering.  {\em Proc. International
Conference of the Chilean Computer Science Society},
IEEE Computer Society Press, 2005, pp. 3-12.}

\ignore{
\bibitem{CB2007} Caniupan, M., Bertossi, L. \newblock
The Consistency Extractor System: Querying Inconsistent Databases using Answer Set Programs.
{\em Proc. of the Scalable Uncertainty Management Conference}, Springer LNCS 4772, 2007, pp. 74-88.
}

\bibitem{monicaJournal08}
Caniupan-Marileo, M. and  Bertossi, L. \
The Consistency Extractor System: Answer Set Programs for Consistent Query Answering in Databases. {\em Data and Knowledge Engineering}, 2010, 69(6):545-572.

\bibitem{CM2005} Chomicki, J. and Marcinkowski, J. Minimal-Change Integrity Maintenance
using Tuple Deletions. {\em Information and Computation}, 2005,
197(1-2):90-121.

\bibitem{chom07}
Chomicki, J. \newblock Consistent Query Answering: Five Easy Pieces.
{\em Proc. International Conference on Database Theory}, Springer LNCS 4353, 2007, pp. 1-17.

\bibitem{DEGV2001} Dantsin, E., Eiter, T., Gottlob, G. and Voronkov, A.
Complexity and Expressive Power of Logic Programming. {\em ACM
Computing Surveys}, 2001, 33(3):374-425.

%\bibitem{DLS1996}
%Doherty, P.,  Lukaszewicz, W. and Szalas, A.
%\newblock A Reduction
%Result for Circumscribed Semi-Horn Formulas. \newblock {\em
%Fundamenta Informaticae}, 28, 3-4, 261- 271, 1996.

\bibitem{DLS1997}
Doherty, P., Lukaszewicz, W. and Szalas, A. \newblock Computing
Circumscription Revisited. A Reduction Algorithm. \newblock {\em
Journal of Automated Reasoning}, 1997, 18(3):297-336.

\bibitem{FI96}
Doherty, P., Lukaszewicz, W. and Szalas, A. \newblock A Reduction Result for Circumscribed Semi-Horn
Formulas. \newblock {\em Fundamenta Informaticae}, 1996, 28(3-4):261-271.

%\bibitem{eiterGottlob95}
%Eiter, T. and Gottlob, G. \newblock
%On the Computational Cost of Disjunctive Logic Programming: Propositional Case.
%{\em Annals of Mathematics and Artificial Intelligence}, 1995,
%15(3-4):257-456.

\bibitem{eiter97}
Eiter, T. and Gottlob, G. \newblock Expressiveness of Stable Model Semantics for
    Disjunctive Logic Programs with Functions. {\em Journal of Logic Programming}, 1997, 33(2):167-178.

\bibitem{eiterGottlob97}
Eiter, T., Gottlob, G. and Mannila, H. \newblock
Disjunctive Datalog.
    {\em ACM Transactions on Database Systems}, 1997,
    22(3):364-418.

    \bibitem{fagin74}
Fagin, R. Generalized First-Order Spectra and Polynomial-Time Recognizable Sets. In {\em Complexity of Computation}, R. Karp (ed.), SIAM-AMS Proceedings 7, 1974, pp. 43-73.

\bibitem{FLL2007} Ferraris, P.,  Lee, J. and Lifschitz,
    V. \newblock A New Perspective on Stable Models. In {\em Proc.
    International Joint Conference on Artificial Intelligence}, 2007, pp. 372-379.

 \bibitem{ExtFerraris}
Ferraris, P., Lee, J. and Lifschitz, V. \
Stable Models and Circumscription. {\em Artificial Intelligence}, 2011, 175(1):236-263.


%\bibitem{FLLPS2001} Franconi, E., Laureti Palma, A., Leone, N., Perri, S. and
%Scarcello, F. \newblock Census Data Repair: a Challenging Application of
%Disjunctive Logic Programming. \emph{Proc. Logic for Programming,
%Artificial Intelligence, and Reasoning (LPAR)}. Springer LNCS
%2250, 2001, pp. 561-578.}

\ignore{\bibitem{FM2003}
Fuxman, A. and Miller, R. \newblock Towards Inconsistency
Management in Data Integration Systems. \newblock {\em Proc.
Workshop on Information Integration on the Web (IIWeb)}, 2003.}

\bibitem{FM2007} Fuxman, A. and Miller, R. \newblock
    First-Order Query Rewriting for Inconsistent
    Databases. {\em J. Computer and Systems Sciences}, 2007, 73(4):610-635.


%\bibitem{GL1988} Gelfond, M., Lifschitz, V. The Stable Model Semantics for Logic
%Programming. In Robert Kowalski and Kenneth Bowen, editors,
%\emph{Proceedings of International Logic Programming Conference
%and Symposium}, pages 1070-1080, 1998.

%\bibitem{gelfondLeone02}
%Gelfond, M. and Leone, N. \newblock
%Logic Programming and Knowledge Representation:
%The A-Prolog Perspective.
%{\em Artificial Intelligence}, 2002, 138(1/2):3-38.

\bibitem{GL1991} Gelfond, M., Lifschitz, V. \newblock Classical Negation in Logic
Programs and Disjunctive Databases. {\em New Generation Computing}, 1991,
9(3/4):365-385.

\bibitem{GGZ2003} Greco, G., Greco, S. and Zumpano, E. \newblock A Logical
Framework for Querying and Repairing Inconsistent Databases.
{\em IEEE Transactions on Knowledge and Data Engineering}, 2003,
15(6):1389-1408.

%\bibitem{GO1992}
%Gabbay, D.M. and Ohlbach, H.J. \newblock Quantifier Elimination in
%Second-Order Predicate Logic. In {\em Proc. KR 92}, 1992, pp.
%425-435.

\bibitem{Paraschos20}
 Koutris, P. and  Wijsen, J. \newblock
First-Order Rewritability in Consistent Query Answering with Respect to Multiple Keys. \newblock Proc. PODS 2020, pp. 113-129.

\bibitem{leone06}
Leone, N., Pfeifer, G., Faber, W., Eiter, T., Gottlob, G., Perri, S. and Scarcello, F. \newblock
The {\em DLV} System for Knowledge Representation and Reasoning. {\em ACM Transactions on Computational Logic}, 2006, 7(3):499-562.



\bibitem{libkin}
Libkin, L. \newblock {\em Elements of Finite Model Theory.} Springer, 2004.

\bibitem{Lifschitz1985}
Lifschitz, V. \newblock Computing Circumscription. \newblock
{\em Proc. International Joint Conference on Artificial Intelligence}, Morgan Kaufmann, 1985, pp. 121-127.

\bibitem{L1994} Lifschitz, V. \newblock Circumscription. In {\em Handbook of
Logic in Artificial Intelligence and Logic Programming}, Vol. 3.
Oxford University Press, 1994, pp.297-352.

\bibitem{L1987} Lloyd, J.W. \newblock {\em Foundations of Logic Programming}. Springer Verlag, 1987.

\bibitem{LoBe07}
Lopatenko, A. and Bertossi, L. \newblock Complexity of Consistent Query Answering in Databases under Cardinality-Based and
Incremental Repair Semantics. {\em Proc. International Conference of Database Theory}, Springer LNCS 4353, 2007, pp. 179-193.

\bibitem{Splitting} Lifschitz, V. and Turner, H.
\newblock Splitting a Logic Program.
\newblock {\em Proc. International Conference on Logic Programming}, MIT Press, 1994, pp. 23-37.

\bibitem{twelve} Lifschitz, V. \newblock Twelve Definitions of Stable Model.
 {\em Proceedings International Conference on Logic Programming}. Springer LNCS 5366, 2008, pp. 37-51.

%\bibitem{LB2006} Lopatenko, A. and  Bertossi, L. Complexity of Consistent
%Query Answering in Databases under Cardinality- Based and
%Incremental Repair Semantics. Technical Report arXiv:cs.DB/0604002
%v1. Posted April 2, 2006.


\bibitem{M1980}
McCarthy, J. \newblock Circumscription - A Form of Non-Monotonic Reasoning.
{\em Artificial Intelligence}, 1980, 13(1-2):27-39.

%\bibitem{M1986}
%McCarthy, J. Applications of Circumscription to Formalizing Common
%Sense Knowledge. {\em Artificial Intelligence}, 1986, 28:89-116.

%\bibitem{NOS1999}
%Nonnengart, A., Ohlbach, H.J. and Szalas, A. \newblock Elimination
%of Predicate Quantifiers. \newblock In `Logic, Language and
%Reasoning. Essays in Honor of Dov Gabbay', Part I, H.J. Ohlbach
%and U. Reyle (eds.), Kluwer, 159- 181, 1999.


%\bibitem{NonnSzalas} Nonnengart, A. and Szalas, A.
%\newblock A Fixpoint
 %   Approach to Second-Order Quantifier
 % Elimination with Applications to Correspondence Theory.
%In {\em Logic at Work: Essays Dedicated to the Memory of Helena
%Rasiowa}, Orlowska, E. (ed.), Springer Physica-Verlag, 1998,
%pp. 307-328.}

%\bibitem{O1996}
%Ohlbach, H.J. SCAN - Elimination of Predicate Quantifiers. In {\em
%Proc. CADE 96}, 1996, pp. 161-165.

\bibitem{przy88}
  Przymusinski, T. \newblock
 On the Declarative Semantics of Deductive Databases and Logic Programs.
 In {\em Foundations of Deductive Databases and Logic Programming}, J. Minker (ed.),
  Morgan Kaufmann Publishers Inc.,  1988, pp. 193-216.

\bibitem{przy91}
  Przymusinski, T. \newblock Stable Semantics for Disjunctive Programs. {\em New Generation Computing},
  1991, 9(3/4):401-424.

\bibitem{R1984}
Reiter, R.
\newblock Towards a Logical Reconstruction of Relational Database Theory.
\newblock In {\em On Conceptual Modelling}, M.L. Brodie, J. Mylopoulos and J.W. Schmidt (eds.), Springer, 1984,
pp. 191-233.

\bibitem{KRHB} Van Hermelen, F., Lifschitz, V. and Porter, B. (eds.) \ {\em Handbook of Knowledge Representation}. Elsevier, 2008.

%\bibitem{W2003} Wijsen, J. Condensed Representation of Database Repairs
%for Consistent Query Answering. \emph{Proc. International
%Conference on Database Theory (ICDT)}, pages 378¨C393.
%Springer-Verlag, LNCS 2572, 2003.

\ignore{\bibitem{W2007} Wijsen, J. \newblock On the Consistent Rewriting of Conjunctive Queries Under Primary Key Constraints.
{\em Proc. International Symposium on Database Programming Languages}, Springer LNCS 4797, 2007, pp. 112-126.
}

\bibitem{wijsenIPL10}
Wijsen, J. \newblock A Remark on the Complexity of Consistent Conjunctive Query Answering
under Primary Key Violations. \newblock
{\em Information Processing Letters}, 2010, 110:950-955.

\bibitem{wijsenIS09}
Wijsen, J. \newblock On the Consistent Rewriting Of Conjunctive Queries under Primary
Key Constraints. \newblock {\em Information Systems}, 2009, 34:578-601.

\bibitem{wijsenSurvey}
Wijsen, J. \newblock A Survey of the Data Complexity of Consistent
Query Answering under Key Constraints. \newblock Proc. FoIKS 2014, LNCS 8367, pp. 62-78.

\bibitem{wijsenSigRec}
Wijsen, J. \newblock
Foundations of Query Answering on Inconsistent Databases. \newblock {\em SIGMOD Record}, 2019, 48(3):6-16.



\end{thebibliography}
\end{document}